\documentclass[12pt,a4paper,tightenlines,nofootinbib]{revtex4}
\usepackage{amssymb}
\usepackage{amsmath}
\usepackage{epsf}
\usepackage{psfrag}

\def\nablaslash{\not{\hbox{\kern-3pt $\nabla$}}}

\begin{document}

\author{Jaume Garriga$^{1}$ and Massimo Porrati$^{2}$}
\affiliation{$^1$ Departament de F{\'\i}sica Fonamental,
Universitat de Barcelona, Diagonal 647, 08028 Barcelona, Spain}
\affiliation{$^2$ Department of Physics, New York University, 4 Washington Pl,
New York NY 10003 (USA)}
\title{Football Shaped Extra Dimensions and the Absence of Self-Tuning}
\date{\today}

\begin{abstract}

There have been some recent claims that brane-worlds of co-dimension two in a 6D bulk with compact extra dimensions may lead to self-tuning of the effective 4D cosmological constant.  Here we show that if a phase transition occurs on a flat brane, so as to change its tension, then the brane will not remain flat. In other words, there is really no self-tuning in such models, which can in fact be understood in four-dimensional terms and are therefore subject to Weinberg's no-go theorem.

\end{abstract}

\maketitle

\section{Introduction and Summary} 

The gravitational field of a flat brane of tension $T$ and co-dimension 2 is characterized by a deficit angle 
\begin{equation}
\delta\phi= M^{2-d}\ T, \label{deficit}
\end{equation}
where $M$ is the reduced Planck mass and $d$ is the bulk spacetime dimension.
Aside from this missing wedge, the brane produces no long range gravitational field. Consequently, certain solutions of the equations of motion containing a flat brane, remain solutions under the replacements $T\to T'$ and $\delta \phi\to \delta\phi'=M^{2-d}\ T'$, while keeping everything else the same. This peculiarity has prompted the search for scenarios where the cosmological constant problem might be ameliorated. The logic is that the contribution of brane fields to the vacuum energy amounts to a renormalization of the brane tension, $\Delta T$, which may cause a corresponding change in the deficit angle $\Delta(\delta\phi)$, but hopefully no effect at all on the world-sheet intrinsic geometry.

Several explicit six dimensional models with compactified extra dimensions have been analyzed, leading to claims that the brane contribution to the effective 4D cosmological constant $\Lambda_{eff}$ is self-tuned away at the classical level.\footnote{Concerns have even been raised that this might be too much of a good thing, since it might prevent inflation from happening. This has been dubbed the "moving target" problem in Ref.\cite{burgess}.} As we shall show in this paper, there is no self-tuning of $\Lambda$ in these models, but just the usual fine-tuning.

We begin in Section 2 with a discussion of the simple non-supersymmetric model presented in Refs. \cite{cagu,navarro1}. This is a flux compactification of 
the 6D Einstein-Maxwell-$\Lambda_6$ Lagrangian on a two sphere. Flat 3-branes of equal tension are added at the poles of the sphere, which have the only effect of removing a deficit angle, leading to a  "foot-ball" shaped internal space. 
Objections to self-tuning in this context have been raised by several authors \cite{navarro2,nilles,lee} on the grounds that the magnetic flux is quantized. The argument presented here is tightly related to these objections \footnote{See also  \cite{graesser} for related discussions.}, but we stress the irrelevance of flux quantization. Self-tuning fails simply because of flux conservation, and even if the flux does not couple to charges (in which case it need not be quantized). Next, in Section 3, we consider the supergravity case, which has been discussed in \cite{ABPQ}. This model has an additional dilaton, whose equation of motion drives the effective four-dimensional vacuum energy to zero. However, when phrased in four-dimensional terms, the effective potential has the form dictated by Weinberg's no-go theorem, and the model fails to self-tune for the same reason it does in the Einstein-Maxwell model.

\section{Einstein-Maxwell}

In this section we shall consider the issue of self-tuning in the context of the original model considered in \cite{cagu,navarro1}. This is based on the six 
dimensional bulk action
\begin{equation}
I= \frac{M^4}{2}\int d^6  x \sqrt{-G} \left( {\cal R}[G] - \frac{1}{2\cdot 2!}F_{[2]}^2 - 2 \Lambda_6 \right). \label{action}
\end{equation}
We shall be interested in solutions of the form
\begin{equation}
ds^2 = e^{-2\psi(x)} g_{\mu\nu}(x) dx^{\mu}dx^{\nu}+ M^{-2} e^{2\psi(x)}d\Sigma^2_{\alpha}.  \label{ansatz}
\end{equation}
Here, 
\begin{equation}
d\Sigma^2_{\alpha}= d\theta^2+\alpha^2 \sin^2\theta d\phi^2,\label{internal}
\end{equation}
is the metric of the internal two-sphere, where we have allowed for a 
deficit angle, 
\begin{equation}
\delta\phi= 2\pi(1-\alpha),
\end{equation}
around the poles (the range of the azimuthal angle is $0\leq\phi< 2\pi$). Moreover, we also allow for a non-vanishing field strength
\begin{equation}
F_{[2]}= b(x)\ {\rm vol}(\Sigma_{\alpha}),\label{twoform}
\end{equation}
where ${\rm vol}(\Sigma_{\alpha})=\alpha\sin\theta d\theta \wedge d\phi$ is the volume form of the internal space (of unit radius).

Branes minimally coupled to the six dimensional metric will also be introduced at the poles, but let us begin by considering the case without branes, since the fine-tuning of the cosmological constant can be formulated already 
at this level.

\subsection{Brane-Less Solution and Fine-Tuning}

 If there are no branes, the internal space is regular, with $\alpha=1$. It can be shown that the equations of motion for the action (\ref{action}) with the ansatz (\ref{ansatz}) and (\ref{twoform})  can be obtained from the following 4D action  (see e.g. \cite{gavi})
\begin{equation}
I=\frac{M_4^2}{2}\int d^4  x \sqrt{-g} \left[ {\cal R}[g] - 4 (\partial\psi)^2-V(\psi) \right], \label{4daction}
\end{equation}
where $M_4^2=4\pi M^2$ and
\begin{equation}
V(\psi) = e^{-2\psi}\left(\frac{b^2}{2} e^{-4 \psi} - 2 K\ M^2 e^{-2 \psi} + 2 \Lambda_6\right). \label{potential}
\end{equation}
The overall factor of $e^{-2\psi}$ in the potential comes from $\sqrt{-G}=M^{-2} e^{-2\psi}\sqrt{g}$. The origin of the three different terms within parenthesis are easy to understand. The first one comes from the $F_{[2]}^2$ term in(\ref{action}). The identity $dF=0$ forces $b={\rm constant}$, and the factors of $e^{-4\psi}$ come from raising the two internal indices in the contraction $F_{ab}F^{ab}$. The second term reflects the curvature of the internal manifold, and the constant $K=+1$ has been introduced for mere bookkeeping ($K=0$ and $K=-1$ would correspond to toroidal and hyperbolic compactifications respectively, which we shall not consider here). The curvature term is inversely proportional to the square of the radius of the internal space, hence the factor $e^{-2\psi}$. The third term in (\ref{potential}) is due to the bulk cosmological constant, and does not have any additional dependence on the radion.

A solution with a flat 4D metric $g_{\mu\nu}=\eta_{\mu\nu}$ is obtained by setting $V(\psi)=V'(\psi)=0$. This requires a relation between the two physical parameters $\Lambda_6$ and the value of the magnetic field $B^2=F_{ab}F^{ab}/2$,
or equivalently, between $\Lambda_6$ and the constant $b$:
\begin{equation}
2 \Lambda_6 = \frac{B^2}{2}= \frac{2M^4}{b^{2}}. \label{fine}
\end{equation}
Imposing (\ref{fine}), the minimum of the radion potential $V(\psi)$ is at
\begin{equation}
R^2 = M^{-2} e^{2\psi} = \frac{b^2}{2 M^4}, \label{radius}
\end{equation}
where $R$ is the physical curvature radius of the internal space. Equation (\ref{fine}) represents the celebrated fine-tuning of the effective four dimensional cosmological constant. 

\subsection{Branes and the issue of self-tuning}

As mentioned above, adding flat 3-branes of {\em identical} tension $T$ at the north and south poles of the internal sphere has the only effect of introducing conical singularities, with deficit angle 
\begin{equation}\delta\phi = 2\pi (1-\alpha) = T/M^4. \label{relation}
\end{equation}
Aside from that, no other changes are needed in the bulk. The requirement of equal brane tension on both poles does {\em not} necessarily mean additional fine-tuning. As mentioned in \cite{cagu}, imposing a $Z_2$ reflection symmetry across the equator of the sphere, both tensions are guaranteed to be identical. Hence, in order to make our discussion clearer, we shall impose this discrete symmetry in what follows. 

The bulk solution has the same fine-tuned values of the parameters (\ref{fine}), and the same curvature radius of the internal manifold (\ref{radius}), as if there were no branes. It is this fact which has led to the hope that the brane contribution to the effective cosmological constant self-tunes itself away. 

That this is not true can be understood by analyzing what would happen when a phase transition suddenly changes the tension of the branes. Clearly, the parameter $\alpha$ characterizing the deficit angle would be different before and after the phase transition. Consider two points $x_1$ and $x_2$ on the brane world-sheet where the tension takes different values. The point $x_1$ is in the old phase, with brane tension $T_1$ and deficit angle characterized by $\alpha_1= 1 - T_1/(2\pi M^4)$, whereas $x_2$ is in the region of the new phase, where the brane tension is $T_2$ and the deficit angle is characterized by $\alpha_2= 1 - T_2/(2\pi M^4)$. 
The tension changes from $T_1$ to $T_2$ across some domain wall of finite width, but we need not be concerned about its detailed profile.\footnote{The boundary between old and new phase could also be space-like, the argument does not really depend on the nature of the interface.} We may simply assume that $x_1$ and $x_2$ are deep inside the old and new phases respectively. Since the field strength is a closed form $dF_{[2]}=0$, the magnetic flux integrated over the internal space must be the same at points $x_1$ and $x_2$:\footnote{This follows from considering a curve $\gamma$ joining the points $x_1$ and $x_2$ and integrating $dF_{[2]}$ on the manifold $\gamma$ times the internal space $\Sigma$.} 
\begin{equation}
\Phi_B=\int_{\Sigma_{\alpha_1}\otimes \{x_1\}} F_{[2]} = 
\int_{\Sigma_{\alpha_2}\otimes\{x_2\}} F_{[2]}.\label{flux}
\end{equation}
 Consequently,
\begin{equation}
\frac{\Phi_B}{4\pi}=\alpha_1 b_1= {\alpha_2} b_2 \equiv b_0.
\label{constancy}
\end{equation}
Clearly, the fine-tuning of $\Lambda_6$, given in Eq. (\ref{fine}), which is 
required for the existence of Minkowski branes, cannot be imposed both for 
$b=b_1$ and $b=b_2$ when $\alpha_1 \neq \alpha_2$. Rewriting Eq. (\ref{fine}) as 
\begin{equation}
\alpha^2=\left(\frac{\Phi_B}{4\pi}\right)^2\frac{\Lambda_6}{M^4}, \label{fine2}
\end{equation}
the left hand side changes at the phase transition, whereas the right hand side cannot.
This shows that self-tuning does not really occur: at least one of the two phases will not have a vanishing 4D effective cosmological constant.

One may try to evade this argument by supposing that there exist two-branes
magnetically charged under the 6D Maxwell field. Then, if the phase transition is accompanied by nucleation of these branes (see~\cite{bt}), 
the magnetic flux changes by an integer multiple of the brane magnetic charge. One may then imagine that this change can compensate the change 
in the 4D scalar potential due to the phase transition. This mechanism is not a self-tuning; it is, indeed, a standard fine-tuning. Moreover, 
most phase transitions, such as the electroweak symmetry breaking, or the (de)confinement phase transiton, are neither due to, nor accompanied by brane nucleation, and so the argument above applies without change.

It is easy to see what will be the quantitative effect of a phase transition. For branes of constant tension at the poles of the sphere, the bulk equations of motion with the ansatz (\ref{ansatz}), (\ref{twoform}) and (\ref{relation}) can still be derived from the four-dimensional action (\ref{4daction}). The only difference is that now
\begin{equation}
M_4^2= 4\pi \alpha M^2,
\end{equation}
since the volume of the internal space depends linearly on $\alpha$. Also,
the potential is given by:
\begin{equation}
V(\psi,\alpha)= e^{-2\psi}\left(\frac{b_0^2}{2\alpha^2} e^{-4 \psi} - 2 K\ M^2 e^{-2 \psi} + 2 \Lambda_6\right). \label{potentialpha}
\end{equation}
Here, we have replaced $b$ by $b_0/\alpha$ according to Eq. (\ref{constancy}). 
When a phase transition occurs on the brane, its effective tension changes by 
$\delta T$, while the position of the minimum $\psi_m$ also shifts by $\delta \psi$. The change in the minimum of the potential is then given by
$$
\delta V = V,_\alpha \delta\alpha,
$$
where we have used that $V,_\psi=0$ at the extrema. After some simple algebra, one finds that
\begin{equation}
\delta V = V \frac{\delta\alpha}{\alpha}+{2\over M_4^2} e^{-4\psi} (2\delta T).
\label{deltav}
\end{equation}
The first term corresponds to bulk contributions to the energy density (including the bulk curvature contribution), which are increased or decreased proportionally to $\alpha$. If the original $V$ is fine-tuned to zero, then this first term is absent.
The second term corresponds to the brane contribution. If the tension changes by $\delta T$, we obtain a contribution $2 \delta T$ since there is a brane at each pole. The factor $2/M_4^2$ is due to the unusual normalization of the potential (the inverse of this factor was pulled out of the integral). Finally, the factor $e^{-4\psi}$ converts the tension per unit physical volume in the six dimensional metric to the tension per unit physical volume as measured by the metric $g_{\mu\nu}$. 

The curvature radius of the internal space is given by $R=M^{-1}e^\psi$. This may stabilize at a size which is much larger than the cut-off scale $M^{-1}$. In this case, it is convenient to express the effective theory around this minimum in terms of the rescaled metric to $\tilde g_{\mu\nu}= e^{-2\psi_0}g_{\mu\nu}$, where $\psi_0$ is the expectation value of $\psi$. Omiting the radion kinetic and mass term, the resulting effective action is $I=(M_P^2/2)\int d^4  x \sqrt{-\tilde g}\ {\cal R}[\tilde g] - \int d^4  x \sqrt{-\tilde g}\ \Lambda_{eff} + ...$, where $M_P^2 = 4\pi \alpha M^4 R_0^2$ is the square of the effective Planck mass, and $\Lambda_{eff}=(M_4^2/2)e^{4\psi_0}V(\psi_0)$. Starting from a vacuum with $\Lambda_{eff}=0$, then after the phase transition we have, from (\ref{deltav}), 
\begin{equation}
\Lambda_{eff}= (M_4^2/2)e^{4\psi_0}\delta V=\ 2\ \delta T.
\label{lambdaeffective}
\end{equation}
Far from being self-tuned, the effective 4D cosmological constant changes at the phase transition by precisely the expected amount, i.e. the physical value of the change in the brane tensions. 

\subsection{A Comment on Flux Quantization}

Arguments similar to the ones presented in the previous subsection have been 
given in several papers \cite{navarro2,nilles,lee}. In these references, however, the failure of self-tuning was attributed to the quantization condition on the flux $\Phi_B$. Indeed, the gauge potential can be chosen to be regular at the north or at the south pole $A_{\pm}= b \alpha [\mp 1 + \cos\theta] d\phi$, but not both simultaneously. The difference at the equator, where both regular expressions overlap, is given by $A_--A_+=2b\alpha d\phi= d(2b\alpha\phi)\equiv d\Lambda(\phi)$. If there are charged fields coupled to $A$ with coupling $g$, single valuedness of the gauge transformation $e^{ig\Lambda(\phi)}$ requires 
$$
b=\frac{n}{2g \alpha}.
$$
It follows from this equation that if the fine-tuning (\ref{fine}) is satisfied for some value of $\alpha$, it will not be satisfied by neighboring values. 
Clearly, this argument is based on the existence of charged fields, which 
couple with strength $g$ to the Maxwell field (a similar topological 
constraint was discussed in \cite{ABPQ}). 

Here, we would like to remark that quantization is quite irrelevant to the issue of self-tuning. The argument presented in the previous subsection is independent on whether there are any sources coupled to the flux, and hence does not rely on quantization. Self-tuning fails because of flux conservation, which is a direct consequence of the identity $dF=0$.

\section{Supergravity Case}

Let us now consider the case of supergravity considered in \cite{ABPQ}, which is
a version of the six dimensional Salam-Sezgin model. The relevant part of the bosonic action is very similar to the Einstein-Maxwell model, except for the presence of an additional dilaton $\varphi$:
\begin{equation}
I= \frac{M^4}{2}\int d^6  x \sqrt{-G} \left( {\cal R}[G] - (\partial \varphi)^2 -
\frac{1}{2\cdot 2!} e^{-\varphi} F_{[2]}^2 - 2 e^{\varphi} \Lambda_6 \right). \label{saction}
\end{equation}
Branes are supposed to couple minimally to the six dimensional metric, without
any factor involving the dilaton.
Just as in the previous section, the equations of motion with the ansatze (\ref{ansatz}) and (\ref{twoform}) can be derived from the 4D action
\begin{equation}
I=\frac{M_4^2}{2}\int d^4  x \sqrt{-g} \left[ {\cal R}[g] - (\partial\varphi)^2- 4 (\partial\psi)^2-V(\psi,\alpha) \right], \label{4daction'}
\end{equation}
where $M_4^2= 4\pi \alpha M^2$ and
\begin{equation}
V(\psi,\alpha)= e^{-2\psi}\left(\frac{b_0^2}{2\alpha^2} e^{-\varphi}e^{-4 \psi} - 2 K\ M^2 e^{-2 \psi} + 2 \Lambda_6\ e^{\varphi}\right). 
\label{potentialpha'}
\end{equation}
A curious fact about this potential is that, regardless of the values of the parameters, the conditions $V,_\varphi=0$ and $V,_\psi=0$ automatically imply that $V=0$. In other words, all extrema of the potential have vanishing vacuum energy: seemingly, self-tuning is at work \footnote{This vanishing of the potential is indeed implied by the general "classical self-tuning" argument presented in Section 3.1 of \cite{ABPQ}}.
Things become less mysterious by expressing the potential in terms of the combinations $\sigma_1=(2\psi+\varphi)$ and $\sigma_2=(2\psi-\varphi)$,
\begin{equation}
V(\psi,\alpha)= e^{-\sigma_2}\left(\frac{b_0^2}{2\alpha^2} e^{-2 \sigma_1} - 2 K\ M^2 e^{- \sigma_1} + 2 \Lambda_6\right). \label{potentialpha''}
\end{equation}
The potential takes the factorized form $V=e^{-\sigma_2}\tilde V(\sigma_1)$, as dictated by Weinberg's no-go theorem \cite{weinberg}. Self-tuning turns out to be just the usual fine-tuning of the parameters in the potential $\tilde V(\sigma_1)$, which is needed for it to vanish at the minimum. 

Note that $\tilde V$ has the same form as in the Einstein-Maxwell case discussed in Section 2. This potential depends on $\alpha$, and cannot self-tune itself to zero because of flux conservation. If a phase transition occurs from Minkowski space so as to reduce the value of the brane tension, the field $\sigma_1$ will roll toward a new minimum where the potential $\tilde V$ is negative.
\footnote{With the same conventions as in the discussion surounding Eq. (\ref{lambdaeffective}), the effective action takes the form
$
I=(M_P^2/2)\int d^4  x \sqrt{-\tilde g}\ \left[{\cal R}[\tilde g]-(\partial\tilde\sigma_2)^2\right] - \int d^4  x \sqrt{-\tilde g}\ V_{eff}(\tilde\sigma_2) + ...
$ 
Here, we have omitted the kinetic and mass terms for the stabilized field $\sigma_1=\sigma_1^{(0)}$, and we have introduced the shifted field $\tilde\sigma_2 = \sigma_2-\sigma_2^{(0)}$, where $\sigma_2^{(0)}$ denotes the value of $\sigma_2$ at any fiducial point $x_0^{\mu}$ well inside the new phase.
The square of the effective Planck mass is given by $M_P^2 = 4\pi \alpha M^4 R_0^2$ , with $R_0$ the physical curvature radius of the internal space at $x_0^{\mu}$, and $V_{eff}(\tilde\sigma_2)=(M_4^2/2)e^{4\psi_0}V(\sigma_1^{(0)},\sigma_2)$. Starting from a vacuum with $V_{eff}=0$, then after the phase transition we have, 
$$
V_{eff}(\tilde\sigma_2)\approx e^{-\tilde\sigma_2}\ (2\ \delta T).
$$
Hence, the energy scale for the runaway potential for $\tilde\sigma_2$ is given by the magnitude of the correction to the brane tension, $\delta T$.
}
For negative $\tilde V$, the field $\sigma_2$ rolls toward ever increasing negative values, corresponding to large values of the dilaton and small values of the size of the internal space. 

Far from leading to a new Minkowski solution, the spacetime after the transition heads toward a big crunch.

\section{More general internal spaces?}

The arguments presented above have used the assumption that the internal space before and after the transition is football shaped, Eq. (\ref{internal}).  Other than that, they take full account of the 6 dimensional dynamics. When the results are expressed in 4D terms, we find that Weinberg's theorem is at work.

In a Minkowski vacuum, the requirement of axisymmetry and $Z_2$ symmetry of the internal space are sufficient to single out the football shaped solution as the only regular solution \cite{gibbons}. If the tensions of the two branes at the poles are different (that is, if we give up $Z_2$ symmetry) then the deformed football solutions for the internal space are also known \cite{gibbons}, but they do nothing to improve the situation described in the previous sections. The tensions at the two poles still have to obey the fine tuning relation 
\begin{equation}
\alpha_N\alpha_S = (\Phi_B/4\pi)^2(\Lambda_6/M^4), 
\end{equation}
which generalizes Eq. (\ref{fine2}). Here, $\Phi_B$ is the conserved flux, the subindices $N$ and $S$ stand for north and south poles, and $\alpha_i= 1 - T_i/(2\pi M^4)$, $(i=N,S)$, where $T_i$ are the tensions. If we start with a Minkowski solution, and then one of the $\alpha_i$ suddenly changes at the phase transition, the
result will not satisfy the above relation (the right hand side of the relation changes, but the left hand side cannot), so we cannot have flat space after the phase transition.

Can the situation improve by considering more general compact internal spaces? Deformations of the internal space are described in the 4D language by a discrete set of 4 dimensional scalars. A finite number of these can be massless, and the rest will be massive. Fields whose mass is much larger than the energy scale of $\Lambda_{eff}$ can be integrated out, and we are left with a 4D low energy theory with a finite number of self interacting scalars (whose potential may or may not have some flat directions). To this low energy theory, Weinberg's theorem applies, and adjustment mechanisms are not possible~\footnote{The only 
way to evade Weinberg's theorem is to modify radically the {\em long-distance}
behavior of gravity so that it can no longer be described by a 4D local theory.
This is the avenue attempted e.g. in~\cite{dgs}.}.

\section{Conclusions}

The question of self-tuning in brane-world models of co-dimension 2 has recently received a great deal of attention. We hope that our comments help clarifying the point that there is no self-tuning in the scenarios with foot-ball shaped extra dimensions. Rather than evading Weinberg's no-go theorem, they nicely illustrate its meaning.

While this paper was being prepared for submission, a very interesting new paper by Vinet and Cline \cite{cline} appeared, with conclusions similar to ours. In that paper, the absence of self-tuning in the Einstein-Maxwell case is exhaustively discussed.

An optimistic hope that the supergravity case may have a more satisfactory behaviour is also expressed in \cite{cline}. Unfortunately, the arguments given in the present paper point in a different direction. If the dilaton is stabilized, the cancellation mechanism is fully analogous to the four-form tuning proposed long ago in the 4D context by Hawking~\cite{hawking} (see also~\cite{bt,pb}). 
If it is not, then any phase transition will 
lead to runaway/big crunch solutions on one of the two different phases.

\section{Acknowledgments}

J.G. is grateful to Georgios Kofinas and Alex Pomarol for interesting discussions. The work of J.G. is partially supported by CICYT grant FPA 2004-04582-C02-02 and DURSI
2001-SGR-0061. M.P. is supported in part by NSF grant PHY-0245068.

\end{document}